\documentclass[pra,onecolumn,superscriptaddress,nofootinbib]{revtex4}
\usepackage[a4paper, left=0.8in, right=0.8in, top=1.0in, bottom=1.0in]{geometry}

\usepackage[utf8]{inputenc}
\usepackage[english]{babel}
\usepackage[T1]{fontenc}
\usepackage{amsmath}
\usepackage{amsfonts}
\usepackage{hyperref}
\allowdisplaybreaks

\usepackage{tikz}
\usepackage{quantikz}
\usetikzlibrary{trees}
\usetikzlibrary{positioning}
\usetikzlibrary{arrows.meta}
\usetikzlibrary{positioning}
\usetikzlibrary{decorations.markings}
\usetikzlibrary{decorations.text}
\usetikzlibrary{matrix}
\usepackage{float}

\usepackage{braket}
\usepackage{natbib}
\usepackage{amsthm}

\usepackage[noend]{algpseudocode}
\usepackage{algorithm,algorithmicx}

\algrenewcommand\alglinenumber[1]{\sf\scriptsize\color{blue}{#1}}
\algrenewcommand\algorithmicrequire{\textbf{Input:}}
\algrenewcommand\algorithmicensure{\textbf{Output:}}

\usepackage{subcaption}

\begin{document}

\title{On the Classical Shadow Nonparametric Bootstrap}

\date{\today}
\author{Eric Ghysels}
\email{eghysels@unc.edu}
\affiliation{University of North Carolina, Chapel Hill, NC, USA} 
\affiliation{Kenan-Flagler Business School, Chapel Hill, NC, USA}
\affiliation{Kenan Institute, Chapel Hill, NC, USA}
\author{Jack Morgan}
\affiliation{University of Chicago, IL, USA}
\affiliation{Kenan Institute, Chapel Hill, NC, USA}

\begin{abstract}
Classical shadows proposed by \cite{huang2020predicting} are an efficient method for constructing an approximate classical description of a quantum state using very few measurements. 
In the paper we propose to enhance classical shadow methods using bootstrap resampling methods. We apply nonparametric bootstrapping to assess the variability and accuracy of estimators by repeatedly sampling with replacement from the observed data, i.e.\ in our case the classical shadow measurements. We show that the bootstrap distributions are very different from the Gaussian approximations. Likewise, the theoretical error bounds are not tight compared to the bootstrap percentiles. Finally, we suggest using resampling tools to make risk assessments.
\end{abstract}

\maketitle

\setcounter{secnumdepth}{0}


An efficient method for constructing an approximate classical description of a quantum state using very few measurements was introduced by \cite{huang2020predicting}. The procedure, called {\it classical shadow}, involves $\log M$ measurements to accurately predict $M$ different functions of the state with high success probability.\footnote{\cite{aaronson2018shadow} and \cite{aaronson2019gentle} coined the term shadow tomography, although according to \cite{aaronson2018shadow} it was actually S.T.~Flammia who originally suggested the term.} To avoid outlier corruption, \cite{huang2020predicting} split the classical shadow up into equally-sized independent sub-samples and estimate sample means for each. A final estimate is obtained via a median-of-means estimation.\footnote{Subsequent papers have expanded the statistical analysis of classical shadows. \cite{yano2023quantum} discuss a Quasi-Maximum Likelihood  Estimator (QMLE) and present its asymptotic distribution using standard Taylor series expansions by analogy with the well established classical case. \cite{lukens2021bayesian}  investigate classical shadows through the lens of Bayesian mean estimation, which they show attains on average significantly lower error in many relevant applications.}

\medskip

The motto of classical shadows is: {\it Measure now, and ask questions later}. Our research suggests to modify this to: {\it Measure now, and ask many more questions later with resampling.}
Namely, we propose to use statistical bootstrap methods applied to classical shadow samples. Bootstrap resampling is a nonparametric statistical method used to assess the variability and accuracy of estimators by repeatedly sampling with replacement from the observed data. Introduced by \cite{Efron1979}, the bootstrap allows for the approximation of the sampling distribution of virtually any statistic without requiring strong parametric assumptions. In practice, a large number of bootstrap samples — each of the same size as the original dataset — are drawn, and the statistic of interest is computed for each. This generates an empirical distribution that can be used to construct confidence intervals, perform hypothesis testing, or estimate standard errors. 

\medskip

The use of the nonparametric bootstrap is particularly relevant because the sampling distribution of classical shadow estimators is typically not Gaussian, especially for small sample sizes or for general observables. Indeed, it is often heavy-tailed and non-symmetric  due to the structure of the measurement process and the non-trivial inversion map used to construct the shadow samples. In some special settings, Gaussian-like behavior can emerge asymptotically through Central Limit Theorem arguments, such as for example with Pauli observables, where the shadow norm is well-behaved, and the Gaussian approximation works reasonably well. However, for general observables, the convergence may be slow, and large deviations remain important.

\medskip

One can think of classical shadows as a Monte Carlo experiment involving random draws of unitary operators. Likewise, bootstrap resampling can also be viewed as a Monte Carlo experiment applied to a given data set. Therefore, bootstrapping $B$ replications of a classical shadow sample of size $N$ is leveraging that data as if we had drawn $B$ classical shadow size-$N$ samples instead of a single one. Hence, bootstrapping greatly enhances the possibilities of predicting and analyzing quantum state properties with synthetic data.

\medskip

It is important to stress that the term bootstrap is used in different ways in the computer science literature and that our paper pertains to what is called statistical bootstrap resampling. In computer science, bootstrapping broadly denotes starting from a minimal core and progressively building a more capable system. The term spans operating system start-up, compiler construction, distributed-system initialization, among others.\footnote{More specifically, for {\it Operating Systems and System Startup} - on bare metal, firmware executes a small bootstrap loader that locates and loads the OS kernel, which then initializes devices and userspace, see e.g.\ \cite{tanenbaum-bos-mos,love-linux-kernel,bovet-cesati-ulk}. {\it Networking bootstrap} - networked hosts historically used BOOTP \cite{rfc951} and later DHCP \cite{rfc2131} to acquire initial configuration (IP address, routers, and boot parameters), a canonical example of ``bootstrapping'' a system state. {\it Compiler Bootstrapping and Self-Hosting} - a compiler is bootstrapped when a minimal compiler, often leading to a self-hosting toolchain; see the standard compiler texts \cite{aho-dragon-2e}. Finally, {\it Distributed Systems and Peer Discovery} - in peer-to-peer overlays, ``bootstrap nodes'' provide the initial entry point to discover other peers. Representative designs include Chord \cite{stoica-2001-chord} and Kademlia \cite{maymounkov-2002-kademlia}, whose lookup procedures require a small, well-known seed to join the network and then expand knowledge iteratively. In the quantum setting, this takes several forms. Most prominent is its use in quantum tomography, see e.g.\ \cite{blume2010optimal}, \cite{faist2016practical}, \cite{gonccalves2018bayesian}, among others. Bootstrap also refers to its use in algorithms such as Variational Quantum Eigensolvers (VQE) or error mitigation protocols, where the term describes an iterative refinement procedure.} To the best of our knowledge the use of nonparametric bootstrap as a statistical procedure applied to classical shadows is novel.

\medskip

One of the key features of prediction with classical shadows is scalability with current and future hardware and the resulting classical shadow can be stored efficiently in classical memory. Our suggestion to augment this with bootstrap capabilities expands the realm of applications of classical shadows from storage and (non)linear point predictions to examining distributional properties, such as tail behavior, all within the confines of standard classical statistical methods. 

\medskip

To that end, a second contribution of our paper is that we advance the notation {\it risk management for quantum circuits} where we go beyond expectation values and suggest to use the distribution of the classical shadow statistics, and in particular their tail behavior, to study the properties of quantum systems. To that end we pair the idea of nonparametric bootstrap of classical shadows with notions familiar to financial risk managers such as Value-at-Risk and Expected Shortfall.

\section{Resampling Classical Shadows \label{sec:resampling}}

Standard tomography suffers from  a curse of dimensionality as  the number of parameters needed to describe a quantum system scales exponentially with the number of its constituents. Inspired by insights from  
\cite{aaronson2018shadow}, classical shadows focus on a less ambitious goal of accurately predicting certain properties instead of the full classical descriptions of a quantum system. Throughout we consider  $n$-qubit systems where the quantum state $\rho$ in $d=2^n$ dimensions is  fixed but unknown.

\subsection{Classical Shadows}

The unknown quantum state $\rho$ is subjected to $i$ = 1, $\ldots,$ $N,$ randomly independently chosen $d\times d$ unitary operators $U_i$ for each $i$ one measurement is performed in the computational basis. More specifically, the unitary operators $U_i,$ $i$ = 1, $\ldots,$ $N$ are drawn from the set of $d\times d$ Haar-random unitaries which are tomographically complete. Each draw applies a random unitary to rotate the state ($\rho \mapsto U_i \rho U^\dagger_i$) and the $n$-bit measurement outcome $|\hat{b}_i\rangle \in \left\{0,1 \right\}^n$ is stored in classical memory bits via the $U^\dagger_i |\hat{b}_i \rangle \! \langle \hat{b}_i| U_i$ description. 
One can think of the average (over both the choice of unitary and the outcome distribution) mapping from $\rho$ to its classical snapshot
$U^\dagger_i |\hat{b}_i\rangle\!\langle\hat{b}_i| U_i$
as a quantum channel $\mathcal{M},$ and apply $\mathcal{M}^{-1}$ to $U^\dagger_i |\hat{b}_i\rangle\!\langle\hat{b}_i| U_i$
in a completely classical post-processing step.
This yields for each $i$ a single classical snapshot $\hat{\rho}_i = \mathcal{M}^{-1} \left( U_i^\dagger |\hat{b}_i \rangle \! \langle \hat{b}_i| U_i\right)$ of the unknown state $\rho.$ 

\medskip

Repeating these measurement $N$ times produces what is called the classical shadow of $\rho,$ namely $\{\hat{\rho}_1, \ldots, \hat{\rho}_N\},$ with $\mathbb{E}_\rho [\hat{\rho}_i]$ = $\rho,$ where $\mathbb{E}_\rho$ is the expectation under the data generating process determined by the unknown quantum state $\rho.$ Hence, the sample average of the classical shadow is one straightforward target of interest. More generally, the targeted predictions include linear functions of the underlying density matrix $\rho,$ or even nonlinear functions of $\rho,$ such as entanglement entropy. Besides expectation values, other examples of linear functionals  include the fidelity with a pure target state, the probability distribution governing the possible outcomes of a measurement, among others. 

\medskip

\begin{figure}[H]
	\begin{center}
		\includegraphics[width=9cm, height=6cm]{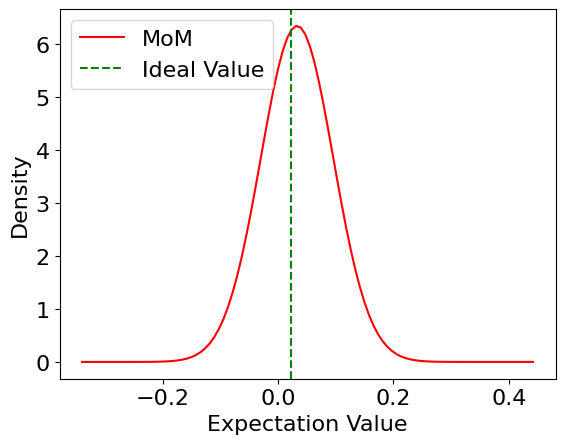}
	\end{center}
	\caption{\label{fig:Gaussian} Gaussian distribution of median-of-mean estimator with $K$ = 10 for $N$ = 1000, for an observable $o_1$ of a 10-qubit circuit involving rotations detailed in the next section, using the estimator defined in equation (\ref{eq:MoM}) and its variance appearing in (\ref{eq:S9}).}
\end{figure}

\noindent Classical shadows are well suited to predict linear functions in the unknown state $\rho:$
\begin{align}
	o_i (\rho) =& \mathrm{tr}(O_i \rho) \quad 1 \leq i \leq M.
	\label{eq:linear_predictions}
\end{align}
where we estimate expectation values $\{o_i\}$ of a set of observables $\{O_i\}$ as:
\begin{equation*}
	\hat{o}_i = \mathrm{tr}(O_i \hat{\rho})
	\quad \Rightarrow \quad
	\mathbb{E}[\hat{o}_i] = \mathrm{tr}(O_i \rho).
\end{equation*}
Hence, the classical shadow provides an unbiased estimator of the linear functional \( \mathrm{tr}(O_i \rho) \).	To avoid outlier corruption, \cite{huang2020predicting} split the classical shadow sample into $K$ equally-sized independent sub-samples of size $N$ and estimate sample means for each, yielding the following median-of-mean estimator:
\begin{equation}
	\label{eq:MoM}
	\hat{o} \equiv	MoM_N = \left[\mathrm{median} \left\{ \mathrm{tr} \left(O_i \hat{\rho}_{(k)} \right)_{k=1}^K \right\}\right]_{i=1}^M,
\end{equation}
for sample estimates $\hat{\rho}_{(k)}$ of subsample $k$ of size $N.$ The fluctuations of shadow estimators depend on a norm that characterizes the chosen measurement primitive:
\begin{equation}
	\|O\|_{\mathrm{shadow}}
	= \max_{\sigma \text{ state}}
	\left(
	\mathbb{E}_{U \sim \mathcal{U}}
	\sum_{b \in \{0,1\}^n}
	\langle b|U \sigma U^\dagger |b\rangle
	\langle b|U\, \mathcal{M}^{-1}(O)\, U^\dagger |b\rangle^2
	\right)^{1/2}.
	\label{eq:S7}
\end{equation}
This is a genuine norm (nonnegative, homogeneous, and obeying the triangle inequality) and encapsulates the statistical expressiveness of the measurement ensemble. Let \( O_0 = O - \tfrac{\mathrm{tr}(O)}{2^n} I \) denote the traceless part of \( O \), and let \( \mathcal{M}^{-1} \) be the adjoint inverse of the measurement map. Then:
\begin{equation}
	\mathrm{Var}[\hat{o}]
	= \mathbb{E}\!\left[(\hat{o} - \mathbb{E}[\hat{o}])^2\right]
	= \mathbb{E}\!\left[
	\bra{\hat{b}} U\, \mathcal{M}^{-1}(O_0)\, U^\dagger \ket{\hat{b}}^2
	\right]
	- \big(\mathrm{tr}(O_0 \rho)\big)^2.
	\label{eq:S9}
\end{equation}
Armed with the median-of-means estimator and its variance we can gauge the statistical properties of the classical shadow estimator.  For a numerical experiment detailed in the next section we provide in Figure \ref{fig:Gaussian} the Gaussian density for the estimator defined in equation (\ref{eq:MoM}) and its variance appearing in (\ref{eq:S9}) for an observable $o_1$ of a 10-qubit circuit involving rotations appearing in Figure \ref{fig:circuit}. Note, that we observe a slight displacement of the distribution compared to the vertical line representing the true expectation value, i.e.\ there is a small sample bias.

\subsection{Bootstrap}

One can think of the sample of independent and identically distributed classical shadow observations $\{\hat{\rho}_1, \hat{\rho}_2, \dots, \hat{\rho}_N\}$ as drawn from an unknown distribution $F$, and let $T_N = T(\hat{\rho}_1, \dots, \hat{\rho}_N)$ be a statistic of interest, such as the sample mean, variance, or any of the functionals (linear or non-linear) mentioned before. The i.i.d.\ bootstrap is a Monte Carlo method for approximating the sampling distribution of $T_N$ using only the observed data.
The procedure can be described by the following steps.
\begin{itemize}
	\item  Construct the empirical distribution function (EDF) $\widehat{F}_N$, defined by: $
	\widehat{F}_N(x)$ = $\frac{1}{N} \sum_{i=1}^N \mathbf{1}_{\{\hat{\rho}_i \leq x\}}.$
	This EDF places equal mass $1/N$ on each observed data point $\hat{\rho}_i$ and can be viewed as the nonparametric maximum likelihood estimator of the true distribution $F$.
	
	\item We consider $b$ = 1, $\ldots,$ $B$ bootstrap samples, typically $B$ is a large number such as $B$ = 10000.
	\item For each $b$ = 1, $\ldots$ we create a sample of size $N$ (called the $b^{th}$ bootstrap sample) $\hat{\rho}_1^{*,b}, \hat{\rho}_2^{*,b}, \dots, \hat{\rho}_N^{*,b}$ which is obtained by drawing $N$ observations {\it with replacement} from the original dataset $\{\hat{\rho}_1, \dots, \hat{\rho}_N\}$. Each $\hat{\rho}_i^{*,b} \sim \widehat{F}_N$ i.i.d.

	\item For each bootstrap sample compute the statistic on the resampled data: $T_N^{*,b}$ = $T(\hat{\rho}_1^{*,b}, \hat{\rho}_2^{*,b}, \dots, \hat{\rho}_N^{*,b}).$
\end{itemize}
Repeat this resampling and computation process independently $B$ times, yielding bootstrap replicates $T_N^{*,1}, T_N^{*,2}, \dots, T_N^{*,B}$.
The empirical distribution of the bootstrap sample $\{T_N^{*,b}\}_{b=1}^B$ approximates the sampling distribution of $T_N$. This distribution can be used to estimate for example the standard error:
$$
\widehat{\text{SE}}(T_N) = \sqrt{ \frac{1}{B - 1} \sum_{b=1}^B \left( T_N^{*,b} - \overline{T_N^*} \right)^2 } \qquad \overline{T_N^*} = \frac{1}{B} \sum_{b=1}^B T_N^{*,b}
$$
or confidence intervals, percentiles, bias-corrected, or studentized intervals can be constructed.\footnote{Biased estimates are:
	$\widehat{\text{Bias}}(T_N)$ = $\overline{T_N^*} - T_N$.} 

\begin{algorithm*}[t]
	{\small
		\begin{algorithmic}[1]
			\caption{{\small Bootstrapping Mean and Median of means prediction  based on a classical shadow.}}
			\label{alg:median_means}
			
			\Statex
			\Function{LinearPredictions}{$O_1,\ldots,O_M,\mathsf{S}(\rho;N),K$}
			
			\State Import $\mathsf{S}(\rho;N) = \left[ \hat{\rho}_1, \ldots, \hat{\rho}_N \right]$
			\Comment Load classical shadow
			
			\For{$b=1$ to $B$} 
			\Comment Start $B$ bootstrap replications
			\State  Draw $N$ observations {\it with replacement} from $\{\hat{\rho}_1, \dots, \hat{\rho}_N\}$ 
			\State Create  {\it bootstrap sample} $\hat{\rho}_1^{(b)}, \hat{\rho}_2^{(b)}, \dots, \hat{\rho}_N^{(b)}$ 
			
			\State Split bootstrap sample $b$ into $K$ equally-sized parts and set
			\Comment Construct $K$ estimators of $\rho$
			\begin{align*}
				\hat{\rho}_{(k)}^{(b)} =& \frac{1}{\lfloor N/K \rfloor} \sum_{i = (k-1) \lfloor N/K \rfloor+1}^{k\lfloor N/K \rfloor} \hat{\rho}_i^{(b)}
			\end{align*}
			\State Compute 
			$
			M_N^{*(b)} = \left[\mathrm{average} \left\{ \mathrm{tr} \left(O_i \hat{\rho}_{i}^{(b)} \right)_{i=1}^N \right\}\right]_{i=1}^M.
			$
			\Comment Mean (M) estimation
			
			\State Compute 
			$
			MoM_N^{*(b)} = \left[\mathrm{median} \left\{ \mathrm{tr} \left(O_i \hat{\rho}_{(k)}^{(b)} \right)_{k=1}^K \right\}\right]_{i=1}^M.
			$
			\Comment Median of means (MoM) estimation
			
			\State $\{M_N^{*(b)}\}_{b=1}^B$ approximates sampling distribution of classical shadow mean estimators
			\State $\{MoM_N^{*(b)}\}_{b=1}^B$ approximates sampling distribution of classical shadow median of mean estimators
			\EndFor
			\EndFunction
		\end{algorithmic}
	}
\end{algorithm*}

There is a well established asymptotic theory underpinning the bootstrap (see e.g.\ \cite{hall1992bootstrap}).
Namely, the i.i.d.\ bootstrap is asymptotically consistent under weak regularity conditions. That is, as $N \to \infty$, the bootstrap distribution converges in probability to the true sampling distribution of $T_N$:
\begin{equation}
	\label{eq:LLN}
	\sup_{x} \left| \mathbb{P}(T_N^* \leq x \mid \hat{\rho}_1, \dots, \hat{\rho}_N) - \mathbb{P}(T_N \leq x) \right| \xrightarrow{P} 0.
\end{equation}
Besides this Law of Large Numbers, the bootstrap Central Limit theorem asserts that:
$$
\sqrt{n}(T_N^* - T_N) \mid \hat{\rho}_1, \dots, \hat{\rho}_N \xrightarrow{d} \mathcal{N}(0, \sigma^2),
$$
and the conditional distribution of the bootstrap replicate $T_N^*$ around the observed value $T_N$ mimics the true sampling distribution of $T_N$ around its population expectation, $\mathbb{E}_\rho[T_N]$ for large $N,$ justifying the use of bootstrap for inference.
This convergence typically holds when $T_N$ is a smooth functional of the empirical distribution, e.g., the mean, variance, quantile (under mild conditions), etc.

\medskip

It is useful to illustrate the bootstrap algorithm where the object of interest is a linear function of an underlying density matrix $\rho,$ e.g.\  the expectation values $\{o_i\}$ of a set of observables $\{O_i\}$:
The fidelity with a pure target state, entanglement witnesses, and the probability distribution governing the possible outcomes of a measurement are all examples that fit this framework.  Algorithm \ref{alg:median_means} describes how to obtain the bootstrap distribution of the mean and the median of means estimators.

\setcounter{equation}{0}

\section{Numerical Experiments \label{sec:numerical}}

In this section we report empirical experiments showcasing the use of the nonparametric bootstrap enhancements of the classical shadows. We use a 10 qubits example of the circuit involving rotation parameters collected in a parameter vector $\theta,$ that is used to prepare a random state. 
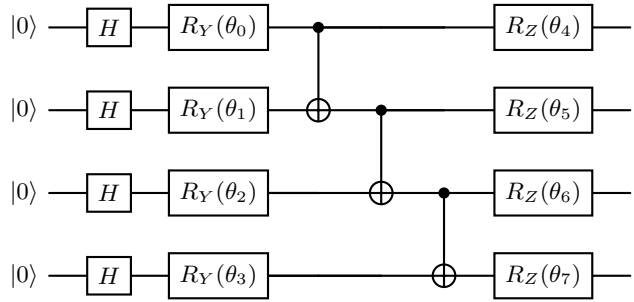
\begin{figure}
	\centering
	\begin{tikzpicture}
		\node{
			\begin{quantikz}
				\lstick{$\ket{0}$} & \gate{H} & \gate{R_Y(\theta_0)} & \ctrl{1} & \qw & \qw & \gate{R_Z(\theta_4)} & \\
				\lstick{$\ket{0}$} & \gate{H} & \gate{R_Y(\theta_1)} & \targ{} & \ctrl{1} & \qw & \gate{R_Z(\theta_5)} & \\
				\lstick{$\ket{0}$} & \gate{H} & \gate{R_Y(\theta_2)} & \qw & \targ{} & \ctrl{1} & \gate{R_Z(\theta_6)} & \\
				\lstick{$\ket{0}$} & \gate{H} & \gate{R_Y(\theta_3)} & \qw & \qw & \targ{} & \gate{R_Z(\theta_7)} & 
			\end{quantikz}
		};
	\end{tikzpicture}
	\caption{\label{fig:circuit} Quantum circuit used to generate classical shadow observables. We use 10 qubit (only 4 displayed) circuit for classical shadows to estimate a set $O$ of 27 observables. }
\end{figure}

We use classical shadows to estimate a set $O$ of 27 observables. $O$ contains the $XX$, $YY$, and $ZZ$ correlations for all adjacent qubits in the circuit, or put another way:
\begin{equation*}
	O = \sum_{i=0}^{n-2} X_i X_{i+1} + Y_i Y_{i+1} + Z_i Z_{i+1}
\end{equation*}
Note that this is not an arithmetic equation. Each term of two Paulis is its own observable. We 'add' each observable in the summation to the set $O$, but we do not create a single observable that is the sum of multiple Paulis. The subscript denotes the qubit which is observed by said Pauli. Figure \ref{fig:circuit} displays the quantum circuit used to generate the classical shadow observables (only 4 displayed in Figure \ref{fig:circuit}).

\begin{figure}[H]
	\begin{center}
		\includegraphics[width=9cm, height=6cm]{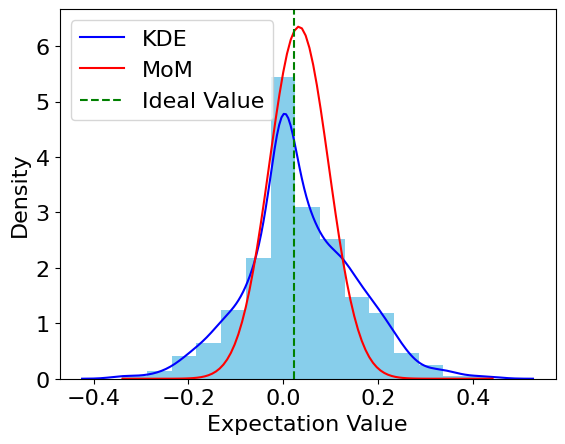}
	\end{center}
	\caption{\label{fig:histogram} The three densities are: (a) the Gaussian appearing in Figure \ref{fig:Gaussian} representing the asymptotic representation, (b) the histogram of the $B$ = 1000 expectation value bootstrap estimates for the median-of-means, and (c) a Kernel density estimate (KDE) smoothed version of the histogram.}
\end{figure}

We implement the classical shadow nonparametric bootstrap introduced in the previous section and display the histogram of the distribution for $B$ = 1000 bootstrap samples each involving a median-of-mean estimator with parameters $K$ = 10 and $N$ = 1000, as appearing in equation (\ref{eq:MoM}).\footnote{All of the code used to generate the figures in this paper is publicly available at \url{https://github.com/jackhmorgan/Classical-Shadow-Nonparametric-Bootstrap}. Our implementation is heavily inspired by the pennylate classical shadows tutorial found here: \url{https://pennylane.ai/qml/demos/tutorial_classical_shadows}.} For illustrative purpose, we focus again on the observable $o_1$ of the 10-qubit circuit (but will discuss other observables as well later). Figure \ref{fig:histogram} displays three distributions, namely: (a) the Gaussian which appeared in Figure \ref{fig:Gaussian} representing the asymptotic representation, (b) the histogram of the $B$ = 1000 expectation value bootstrap estimates for the median-of-means, and (c) a Kernel density estimate (KDE) smoothed version of the histogram. The bootstrap distribution, either depicted by the histogram or by KDE, is non-Gaussian. We notice (a) heavy tails, (b) some hint of asymmetry and (c) similar to the Gaussian approximation a minor small sample bias.  Figure \ref{fig:histogram} tells us that aside from the modest small sample bias, present both in the Gaussian and bootstrap distributions, we have a serious problem of the approximating the tails with the Gaussian density.

\medskip

To shed further light on the inadequacy of the asymptotic distribution we examine the error bounds on the estimation error suggested by \cite{huang2020predicting}. They show that to achieve accuracy \( \varepsilon \) and confidence \( 1 - \delta \)
for \( M \) observables \( \{ O_i \}_{i=1}^M \),
the parameters $K$ and $N$ for the median-of-mean estimator are chosen as
\begin{equation}
	K = 2 \log\!\left( \frac{2M}{\delta} \right),
	\qquad
	N = \frac{34}{\varepsilon^2}
	\max_{1 \le i \le M}
	\left\|
	O_i - \frac{\mathrm{tr}(O_i)}{2^n} I
	\right\|_{\mathrm{shadow}}^2.
	\label{eq:S13}
\end{equation}
This yields an error bound to achieve the accuracy $\varepsilon$ for the estimator $\hat{o}$ in equation (\ref{eq:MoM} )with probability at least \( 1 - \delta \), namely:
\begin{equation}
	\big| \hat{o}_i - \mathrm{tr}(O_i \rho) \big|
	\le \varepsilon,
	\quad
	\forall\, 1 \le i \le M.
	\label{eq:S14}
\end{equation}
This ensures uniform accuracy over all \( M \) observables
using \( N K \) independent classical shadows.

\medskip

\begin{figure}[H]
	\begin{center}
		\includegraphics[width=9cm, height=6cm]{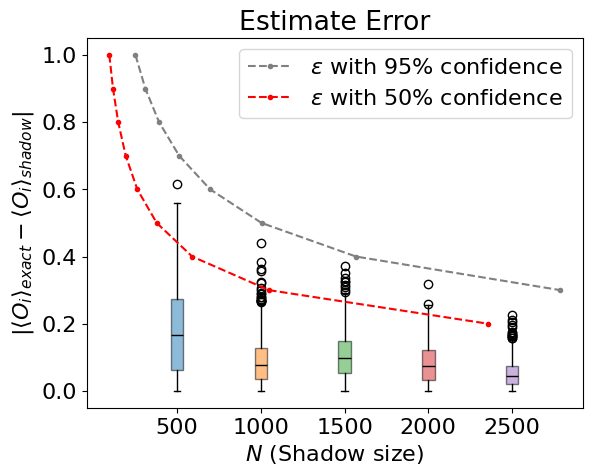}
	\end{center}
	\caption{\label{fig:errorbound} Theoretical error bound implied by equation and bootstrap distributions with $B$ = 1000 with $N$ appearing on the x-axis.}
\end{figure}

Figure \ref{fig:errorbound} displays the error bounds for $\varepsilon$ = $0.95$ and $0.50,$ with $N$ = 500, 1000, 1500, 2000, 2500 and 3000 and $B$ = 1000. We clearly see from the box-plotted bootstrap distributions that the (asymptotic) error bounds are not tight. Hence, we should rely on the bootstrap distibutions to assess the tail behavior. Moreover, the figure also shows that the small sample bias diminishes as we increase $N,$ as is expected.

\setcounter{equation}{0}

\section{Risk Management for Quantum Circuits \label{sec:FaR}}

In this section we suggest concepts that are commonly used in  financial risk management for the purpose of assessing extreme outcomes of quantum circuits.  In financial risk, we only care about the left tail, i.e.\ the downside risk. In some applications involving quantum circuits it might be more appropriate to look at both tails, as suggested by Figure \ref{fig:histogram}. We will focus here on the left tail as it closely follows the financial risk literature.

\medskip

Two measures pertaining to the tail behavior stand out in financial risk management. We start with their definitions and then modify them to the application at hand. The first is Value-at-Risk (V@R), a widely used risk measure that quantifies the maximum potential loss of a portfolio over a specified time horizon at a given confidence level. Formally, for $T_N^*$ (in financial applications typically a loss $L$ on a portfolio of assets - in our case the measured expectation value $T_N$), the $\alpha$-V@R is the $\alpha$-quantile of its distribution:
\[
\text{V@R}_\alpha(T_N) = \inf \{ t \in \mathbb{R} : \mathbb{P}(T_N \leq t) \geq \alpha \}.
\]
At confidence level $\alpha = 0.95$, the 95\%-V@R is the loss threshold such that only 5\% of losses are expected to exceed it. V@R is intuitive and easy to compute, which explains its regulatory adoption, see \cite{jorion2007value} among others for further discussion. V@R has its limitations as it is not subadditive, meaning it does not always satisfy the coherence property of risk measures. As a result, diversification benefits can be underestimated. 

\medskip

A second commonly used measure is Expected Shortfall (ES) (also called Conditional V@R or CV@R). It addresses V@R’s limitations by considering the average loss beyond the V@R level. For a loss variable $L$,
\[
\text{ES}_\alpha(T_N) = \mathbb{E}[T_N \mid T_N \geq \text{V@R}_\alpha(T_N)].
\]
ES gives the expected size of losses in the worst $(1-\alpha)\%$ tail. For example, at $\alpha = 0.95$, ES represents the mean loss among the worst 5\% of outcomes. ES is a coherent risk measure - as discussed by \cite{artzner1999coherent} - satisfying subadditivity, monotonicity, translation invariance, and positive homogeneity. 

\medskip

\begin{table}[H]
	\centering
	\captionsetup[subtable]{justification=centering}
	\begin{subtable}[t]{0.9\textwidth}
		\centering
			\begin{tabular}{lrrcrrccrrcrr}
				& \multicolumn{5}{c}{Bootstrap} & & & \multicolumn{5}{c}{Gaussian} \\ \cline{2-6} \cline{9-13} 	& & & & & & & & & & & & \\
					& \multicolumn{2}{c}{\quad 5\%} & & \multicolumn{2}{c}{\quad 10\%} & & & \multicolumn{2}{c}{\quad 5\%} & & \multicolumn{2}{c}{\quad 10\%} \\
				& \multicolumn{1}{c}{EV@R} & \multicolumn{1}{c}{ES} & & EV@R & \multicolumn{1}{c}{ES} & & & EV@R & \multicolumn{1}{c}{ES} & & EV@R & \multicolumn{1}{c}{ES} \\
				& & & & & & & & & & & & \\
				& -0.1533 & -0.2072 & & -0.1055 & -0.1676 & & & -0.0.0703 & -0.0966 & & -0.0475 & -0.0772 \\
			\end{tabular}
		\caption{Expectation Value at Risk (EV@R) and Expected shortfall (ES) of the measurement based on the distributions appearing in Figure~\ref{fig:histogram}.}
	\end{subtable}
	
	\vspace{1em}
	
	\begin{subtable}[t]{0.9\textwidth}
		\centering
			\begin{tabular}{lrrcrr}
				& \multicolumn{5}{c}{|Bootstrap - Gaussian|}\\ \cline{2-6} & & & & & \\
							& \multicolumn{2}{c}{\quad 5\%} & & \multicolumn{2}{c}{\quad 10\%}  \\
				& EV@R & \multicolumn{1}{c}{ES} & & EV@R & \multicolumn{1}{c}{ES}  \\	
				& & & & & \\
				mean & 0.0696 & 0.0887 & & 0.0537 & 0.0749 \\
				std & 0.0292 & 0.0406 & & 0.0210 & 0.0321 \\
			\end{tabular}
	\caption{Average and standard deviation - computed across all 27 observables of quantum circuit appearing in Figure \ref{fig:circuit} - of the absolute value of the difference between Value-at-Risk and Expected Shortfall, denoted |Bootstrap - Gaussian|, of all operators in $O$ when calculated with the non-parametric bootstrap versus the median-of-means distribution.}
	\end{subtable}
	
	\caption{Comparison between bootstrap and Gaussian Value-at-Risk and Expected Shortfall estimates.}
	\label{tab:far}
\end{table}
\medskip

We suggest to think beyond computing mean values (or robust median-of-mean) and adopt the nonparametric bootstrap for characterizing the tail behavior. While we focus on expectation values (EV), the ideas go beyond this particular application and applies more broadly to classical shadow nonparametric bootstrap output applied to fidelity of circuits, error correction schemes, etc. 
In the context of the current paper, we define EV-at-Risk, or EV@R, as the $\alpha$ level expectation value threshold of potential poor outcomes and the associated expected shortfall. 
In panel (a) of Table \ref{tab:far}  we report the Gaussian approximation and bootstrap distribution 5\% and 10\% EV@R and ES for the $\hat{o}_1$ observable.  Across both levels of $\alpha$ we see some large discrepancies between the Gaussian and bootstrap values, confirming the findings report for $\hat{o}_1$ in Figure \ref{fig:histogram}. Take the 5\% EV@R as an example. From Panel (a) we know that with the bootstrap it is $-0.1533$ and with the Gaussian it is $-0.0703.$ That is a 50\% error in the tail estimate. In panel (b) of Table \ref{tab:far} we report the average and standard deviation - computed across all 27 observables of the quantum circuit appearing in Figure \ref{fig:circuit} - of the absolute value of the difference between Value-at-Risk and Expected Shortfall, denoted |Bootstrap - Gaussian|, when calculated with the non-parametric bootstrap versus the median-of-means distribution. On average, across all observables the Gaussian and bootstrap 5\% EV@R is off by roughly $0.07$ with a standard deviation of $0.03.$ Hence, although we focused mostly on $\hat{o}_1,$ the finding apply across all observables.

\medskip

A word of caution is in order regarding bootstrapping tail probabilities, for central regions, the bootstrap is first-order valid, i.e.:
$\sup_t \big| P(T_N \le t) - P^*(T_N^* \le t) \big| \to 0,$ under standard smoothness and moment conditions \citep{hall1992bootstrap, vanderVaart1996}.  
However, this validity does not automatically extend to tail regions, where approximation errors can grow rapidly. Namely, the bootstrap accuracy depends on the tail regime: (a) for moderate deviations, i.e.\ $t = o(N^{1/6})$, one often has $P(T_N > t)$ $\approx$ $P^*(T_N^* > t) + o_P(1),$ see \citep{hall1992bootstrap}, and (b) for large deviations, when $t = O(\sqrt{N})$, the bootstrap typically fails to reproduce the correct exponential tail rate: $\frac{1}{N} \log P(T_N > t\sqrt{N})$ $\not\approx$ $\frac{1}{N} \log P^*(T_N^* > t\sqrt{N}),$ as discussed by \citep{arcones1992bootstrap} and \citep{spokoiny2012parametric}.
There are several refined bootstrap techniques improve tail estimation, including (a) Studentized bootstrap, (b) Edgeworth-corrected bootstrap (\citep{lahiri2003resampling}), (c) $m$-out-of-$n$ bootstrap or subsampling, see \citep{politis1999subsampling}, among others. Moreover, recent approaches (e.g., \citep{spokoiny2012parametric, cherno2017central}) analyze the finite-sample concentration behavior of bootstrap approximations. We leave these topics for future research.

\medskip

\section{Conclusions \label{sec:conclusion}}

We proposed to use statistical bootstrap methods applied to classical shadow samples.  Classical shadows can be viewed as Monte Carlo experiments involving random draws of unitary operators and  bootstrap resampling are Monte Carlo experiments applied to a given data set. Our suggestion to apply bootstrap capabilities expands to a wide range of applications of classical shadows.

\medskip

While our focus was on a single application, the bootstrap resampling methods apply to many other settings to which classical shadows have been applied. A non-exhaustive list of such applications where our resampling methods can provide new insights include (a) estimating local observables, entropies, and correlators, (b) entanglement detection and quantification, (c) estimation of out-of-time-ordered correlators, (d) Hamiltonian learning,  (e) error mitigation, such as shadow distillation, probabilistic error cancellation, and symmetry-adjusted shadows, among others. But there are also many applications where classical shadows still need to make inroads and the resampling methods make the applications even more appealing. \cite{ghisoni2024shadow} suggest a shadow quantum linear solver type HHL algorithm. \cite{morgan2025enhanced} and \cite{ghysels2024quantum} explore quantum solutions to dynamic asset pricing models with decision-theoretic foundations of observables. The measurement operators considered include those which emphasize tail behavior of asset returns, and resampling would enhance such approaches.

\medskip 

The resampling methods introduced in this paper allows researchers to go beyond the computation of point estimates and asymptotic approximations of tail behavior. We take advantage of the many existing resampling statistical methods to vastly expand the scope of measurement first and asking questions later. 

\medskip

We also advocate to think in terms of risk management when studying the properties of quantum circuits. As in financial risk management, we suggest to ask questions like: what are the worst 5\% tail risks that can happen when looking at a particular circuits and use better tools to answer these questions. Such risks are driven by the design fundamentals of the circuit and the measurements being applied.

\newpage

\bibliographystyle{abbrv}
\bibliography{literature}

\end{document}